\documentclass[3p,times,twocolumn]{elsarticle}
 \biboptions{comma,sort&compress}
 
\usepackage{graphicx}
\usepackage{amsmath}
\usepackage{here}
\usepackage{ecrc}

\volume{00}

\firstpage{1}

\journalname{Nuclear and Particle Physics Proceedings}
\runauth{Wei Chen}

\jid{nppp}
\jnltitlelogo{Nuclear and Particle Physics Proceedings}

\usepackage{amssymb}

\usepackage[figuresright]{rotating}

\begin{document}

\begin{frontmatter}

\title{
%
Searching for fully-heavy tetraquark states in QCD moment sum rules\,$^*$} 
 
 \cortext[cor0]{Mini-Review talk presented at QCD21 - 36 years later, 24th International Conference in QCD (5-9/7/2021,
  Montpellier - FR). }
  
 \author[label1]{Wei Chen\fnref{fn1}}
 \fntext[fn1]{Speaker, Corresponding author.}
\ead{chenwei29@mail.sysu.edu.cn}
\address[label1]{School of Physics, Sun Yat-Sen University, Guangzhou 510275, China
}
\author[label1]{Qi-Nan Wang}
\author[label1]{Zi-Yan Yang}
 \author[label2]{Hua-Xing Chen
 \corref{cor1} 
 }
\address[label2]{School of Physics, Southeast University, Nanjing 210094, China
}
 \author[label3]{Xiang Liu
 }
\address[label3]{School of Physical Science and Technology, Lanzhou University, Lanzhou 730000, China
}
 \author[label4]{T.~G.~Steele
 }
\address[label4]{Department of Physics and Engineering Physics, University of Saskatchewan, Saskatoon, Saskatchewan, S7N 5E2, Canada
}
 \author[label5]{Shi-Lin Zhu
 }
\address[label5]{School of Physics and Center of High Energy Physics, Peking University, Beijing 100871, China
}

\pagestyle{myheadings}
\markright{ }
\begin{abstract}
\noindent
In this talk, we briefly report the investigations of the mass spectra for the $cc\bar c\bar c, bb\bar b\bar b$, $bc\bar b\bar c$ and $cc\bar b\bar b$ tetraquark states by using the QCD moment sum rule method. The calculations for the fully-charm $cc\bar c\bar c$ tetraquarks have been successfully predicted the existence of di-$J/\psi$ resonances including $X(6900)$ in LHCb's observation. The quantum numbers for these resonance structures are also suggested. More efforts are still needed in both theoretical and experimental aspects to study the properties of these fully-heavy tetraquark states. They may be observed at facilities such as LHCb, CMS and RHIC in the future.
 
\end{abstract}
\end{frontmatter}
\section{Introduction}
Plenty of charmoniumlike exotic states including the so called XYZ and $P_{c}$ states have been observed after 2003. Since they can not be fitted into the conventional quark model, many exotic hadron configurations have 
been proposed to understand their inner structure, such as hadron molecules, compact multiquarks, hybrid mesons and so on~\cite{2016-Chen-p1-121,2017-Ali-p123-198,2018-Guo-p15004-15004,2019-Liu-p237-320,2020-Brambilla-p1-154}. 

Very recently, the LHCb Collaboration reported a narrow resonance $X(6900)$ with the significance more than $5\sigma$, a broad structure ranging from 6.2 to 6.8 GeV and a hint for another structure around 7.2 GeV in the di-$J/\psi$ mass spectrum~\cite{LHCb:2020bwg}. These structures are immediately considered as good candidates of compact fully-charm tetraquark states since the absence of light quarks~\cite{Maiani:2020pur,Chao:2020dml}.
Actually, there have been already some theoretical investigations in fully-heavy tetraquark states before LHCb's observation~\cite{Chen:2016jxd,Anwar:2017toa,Esposito:2018cwh,Hughes:2017xie,Karliner:2016zzc,Wu:2016vtq,Richard:2017vry,Bai:2016int,Chen:2019dvd,Debastiani:2017msn}. After LHCb reported $X(6900)$ and another two potential exotic structures, many theoretical researches are raised to study the fully-charm $cc\bar c\bar c$ tetraquarks for their mass spectra~\cite{Albuquerque:2020hio,Giron:2020wpx,Gordillo:2020sgc,Guo:2020pvt,Jin:2020jfc,Karliner:2020dta,Ke:2021iyh,Li:2021ygk,Liang:2021fzr,Liu:2019zuc,liu:2020eha,Pal:2021gkr,Sonnenschein:2020nwn,Wan:2020fsk,Wang:2018poa,Wang:2019rdo,Wang:2020ols,Wang:2021kfv,Weng:2020jao,Yang:2020rih,Yang:2020wkh,Zhang:2020xtb,Zhao:2020zjh,Zhao:2020nwy,Zhu:2020xni,Cao:2020gul,Mutuk:2021hmi,Yang:2021hrb}, their production mechanisms~\cite{Huang:2021vtb,Feng:2020riv,Wang:2020gmd,Feng:2020qee,Maciula:2020wri,Goncalves:2021ytq,Ma:2020kwb,Wang:2020tpt,Zhao:2020nwy,Zhu:2020sn,Gong:2020bmg} and their decay properties~\cite{Guo:2020pvt,Lu:2020cns,Li:2019uch,Chen:2020xwe,Becchi:2020uvq,Sonnenschein:2020nwn}. 

In this talk, I shall present recent progresses on the calculations of the mass spectra for the fully-heavy $cc\bar c\bar c, bb\bar b\bar b$, $bc\bar b\bar c$ and $cc\bar b\bar b$ tetraquark states by using the method of QCD moment sum rules.

\section{Formalism of QCD sum rules}
To introduce the QCD sum rule method~\cite{Reinders:1984sr,Shifman:1978bx}, we start from the two-point correlation functions for the scalar, vector and tensor interpolating currents $J(x), J_\mu(x)$ and $J_{\mu\nu}(x)$ 
\begin{align}
\Pi\left(p^{2}\right)&= i \int d^{4} x e^{i p \cdot x}\left\langle 0\left|T\left[J(x) J^{\dagger}(0)\right]\right| 0\right\rangle\, ,
\\
 \Pi_{\mu \nu}\left(p^{2}\right) &= i \int d^{4} x e^{i p \cdot x}\left\langle 0\left|T\left[J_{\mu}(x) J_{\nu}^{\dagger}(0)\right]\right| 0\right\rangle\, ,
\end{align}

\begin{align}
 \Pi_{\mu \nu,\,\rho \sigma}\left(p^{2}\right) &=i \int d^{4} x e^{i p \cdot x}\left\langle 0\left|T\left[J_{\mu\nu}(x) J_{\rho\sigma}^{\dagger}(0)\right]\right| 0\right\rangle
\\ \nonumber
&=\left(\eta_{\mu\rho}\eta_{\nu\sigma}+\eta_{\mu\sigma}\eta_{\nu\rho}-\frac{2}{3}\eta_{\mu\nu}\eta_{\rho\sigma}\right) \Pi_{2}\left(p^{2}\right)
\\&+\cdots \, ,
\end{align}
where
\begin{equation}\
\eta_{\mu\nu}=\frac{p_{\mu} p_{\nu}}{p^{2}}-g_{\mu \nu}\, ,
\end{equation} 
and the invariant functions $\Pi_{2}\left(p^{2}\right)$, $\Pi_{1}\left(p^{2}\right)$, $\Pi_{0}\left(p^{2}\right)$ relate to the spin-2, spin-1 and spin-0 intermediate states, respectively. To study the fully-heavy tetraquark systems, we use the  interpolating currents constructed for the $cc\bar c\bar c, bb\bar b\bar b, bc\bar b\bar c, bb\bar c\bar c$ tetraquarks 
in Refs.~\cite{Chen:2016jxd,Yang:2021zrc,Wang:2021taf} to calculate the above correlation functions.

The invariant functions can be described by the dispersion relation at the hadronic level 
\begin{equation}
\begin{aligned}
\Pi\left(p^{2}\right)&=\frac{\left(p^{2}\right)^{N}}{\pi} \int_{s<}^{\infty} \frac{\operatorname{Im} \Pi(s)}{s^{N}\left(s-p^{2}-i \epsilon\right)} d s
\\ &+\sum_{n=0}^{N-1} b_{n}\left(p^{2}\right)^{n}\, ,
\end{aligned}
\label{Cor-Spe}
\end{equation}
where $b_n$ is the subtraction constant. The imaginary part of the correlation function can be written as the sum 
over a series of $\delta$ functions
\begin{align}
\text{Im}\Pi(s)&=\pi\sum_n\delta(s-m_n^2)\langle0|J |n\rangle\langle n|J^+|0\rangle
\\
&=\pi f_{H}^{2}\delta(s-m_{H}^{2})+\cdots\, ,
\end{align}
in which the ``pole plus continuum'' ansatz is adopted in the last step, and the ``$\cdots$'' represents the contributions from the higher excited states and the continuum. The parameters $f_{H}$ and $m_{H}$ are the coupling constant and mass of the lowest-lying resonance 
\begin{equation}
\begin{aligned}
\langle 0|J| H\rangle &= f_{H}\, , \\
\left\langle 0\left|J_{\mu}\right| H\right\rangle &= f_{H} \epsilon_{\mu}\, , \\
\left\langle 0\left|J_{\mu\nu}\right| H\right\rangle &= f_{H} \epsilon_{\mu\nu}\, ,
 \end{aligned}
\end{equation}
with the polarization vector $\epsilon_{\mu}$ and polarization tensor $\epsilon_{\mu\nu}$.

At the quark-gluonic level, the invariant function $\Pi(p^{2})$ can be evaluated via the operator product expansion (OPE) method, and expressed as the function of various QCD condensates, the heavy quark masses and QCD coupling $\alpha_s$. For the fully-heavy tetraquark systems, we calculate the correlation functions up to gluon condensate since the higher dimensional nonperturbative terms such the tri-gluon condensate are very small to be neglected. We shall not show the expressions of these invariant functions here since they are very lengthy 
and complicated.

We introduce the following moment as derivatives of the invariant function $\Pi(p^{2})$ in Euclidean region $Q^{2}=-q^{2}>0$ 
\begin{align}
M_{n}\left(Q_{0}^{2}\right)&=\left.\frac{1}{n !}\left(-\frac{d}{d Q^{2}}\right)^{n} \Pi\left(Q^{2}\right)\right|_{Q^{2}=Q_{0}^{2}}
\\ 
&=\int_{s<}^{\infty} \frac{\rho(s)}{\left(s+Q_{0}^{2}\right)^{n+1}} d s\, .
\end{align}
Applying to Eq.(\ref{Cor-Spe}), the moment can be expressed as 
\begin{equation}
M_{n}\left(Q_{0}^{2}\right)=\frac{f_{H}^{2}}{\left(m_{H}^{2}+Q_{0}^{2}\right)^{n+1}}\left[1+\delta_{n}\left(Q_{0}^{2}\right)\right]\,, \label{moments}
\end{equation}
where $\delta_{n}(Q_{0}^{2})$ contains the contributions from higher excited states and continuum. It will tend to zero as $n$ going to infinity for a specific value of $Q_{0}^{2}$. Defining the following ratio 
\begin{align}
r\left(n, Q_{0}^{2}\right) &\equiv \frac{M_{n}\left(Q_{0}^{2}\right)}{M_{n+1}\left(Q_{0}^{2}\right)}
\\
&=\left(m_{H}^{2}+Q_{0}^{2}\right) \frac{1+\delta_{n}\left(Q_{0}^{2}\right)}{1+\delta_{n+1}\left(Q_{0}^{2}\right)}\, ,
\end{align}
where the relation $\delta_{n}(Q_{0}^{2})\approx \delta_{n+1}(Q_{0}^{2})$ will be achieved when $n$ is large enough. Finally, the hadron mass of the lowest-lying hadron state can be extracted as 
\begin{equation}
m_{H}(n,\, Q_{0}^{2})=\sqrt{r\left(n, Q_{0}^{2}\right)-Q_{0}^{2}}\, , \label{hadronmass}
\end{equation}
which is the function of $n$ and $Q_{0}^{2}$.

\section{Moment QCD sum rules and numerical results for the $cc\bar c\bar c, bb\bar b\bar b, bc\bar b\bar c, bb\bar c\bar c$ tetraquark systems}
We use the following values of heavy quark masses and gluon condensate to perform numerical analyses~\cite{Nielsen:2009uh,Narison:2018nbv,ParticleDataGroup:2020ssz}
\begin{equation}
\begin{array}{l}
{m_{c}\left(m_{c}\right)=(1.27 _{-0.02}^{+0.02}) \mathrm{GeV}}\, , \vspace{1ex} \\
{m_{b}\left(m_{b}\right)=(4.18 _{-0.02}^{+0.03}) \mathrm{GeV}}\, , \vspace{1ex} \\
{\left\langle g_{s}^{2} G G\right\rangle= (0.88\pm0.25) \mathrm{GeV}^{4}}\, .
\end{array}
\end{equation}

\begin{figure}[hbt]
\begin{center}
\scalebox{0.43}{\includegraphics{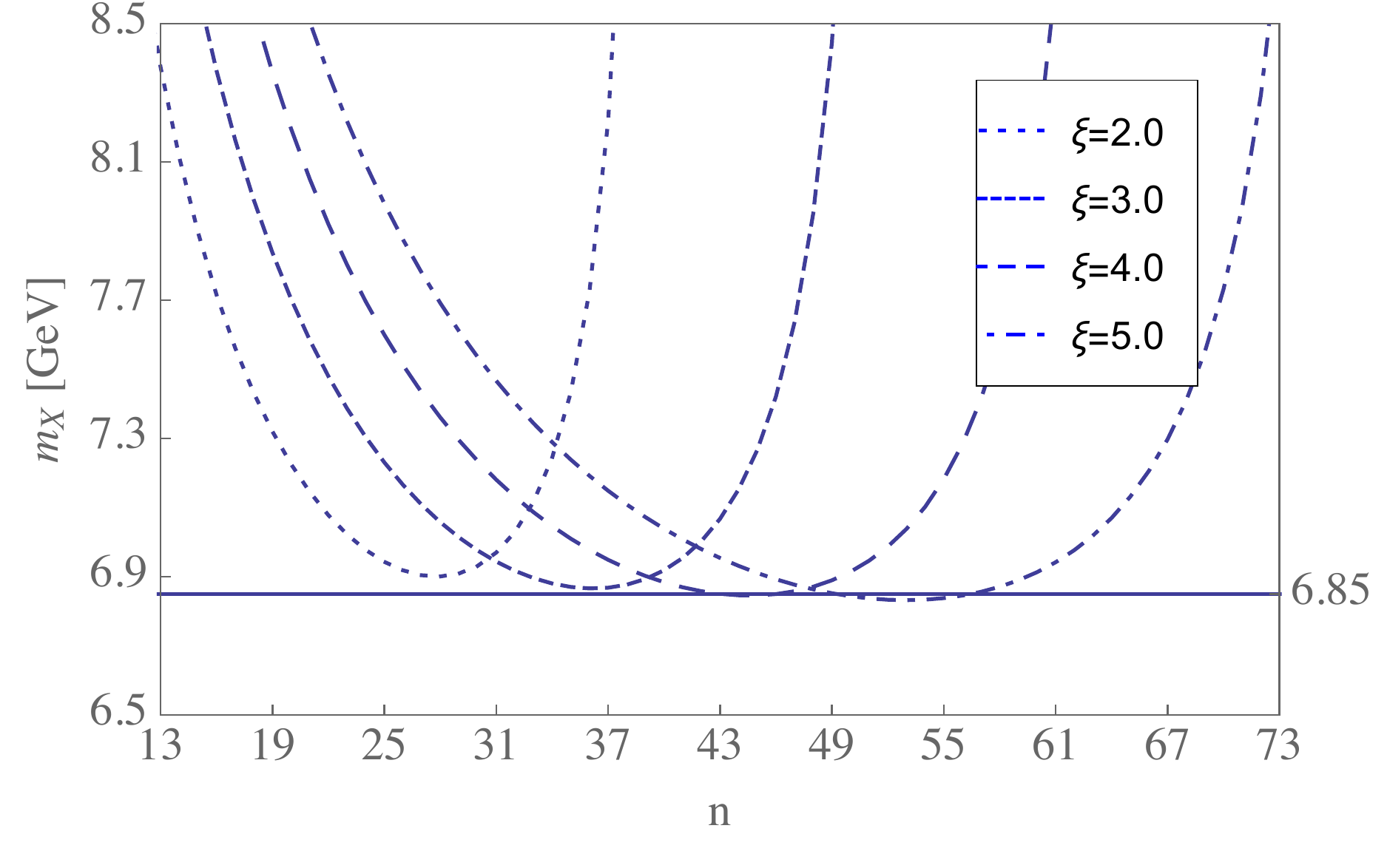}}
\caption{Hadron mass for the fully-charm $cc\bar c\bar c$ tetraquark state with $J^{PC}=0^{-+}$.} \label{cccc0-+}
\end{center}
\end{figure}

As shown in Eq.~\eqref{hadronmass}, there are two important parameters $n$ and $Q_{0}^{2}$ in the hadron mass.
One needs to carefully choose the suitable working regions for these two parameters. For simplicity, we define 
$\xi=Q^2_0/16m_c^2, \xi=Q^2_0/m_b^2, \xi=Q^2_0/(m_b^2+m_c^2)$ for $cc\bar c\bar c, bb\bar b\bar b$ and $bc\bar b\bar c/bb\bar c\bar c$ tetraquark systems, respectively.
These two parameters, $n$ and $\xi$, are correlated with each other:
\begin{enumerate}
\item A larger value of $\xi$ results in slower convergence of $\delta_{n}(Q_0^2)$, which can be compensated by taking higher derivative $n$ of $\Pi(q^2)$ for the lowest lying resonance to dominate.
\item However, a large value of $n$ means moving further away from the asymptotically free region. The OPE
convergence would also become bad for large value of $n$.
\end{enumerate}

\begin{table}[h!]
\renewcommand\arraystretch{1.3} 
\caption{Masses for the fully-charm $cc\bar c\bar c$ and fully-bottom $bb\bar b\bar b$
tetraquark states with various $J^{PC}$ quantum numbers. \label{mass:QQQQ}}
\vspace{6pt}
\begin{tabular}{cccc}
\hline\hline
~~$J^{PC}$ ~~& Currents &~~ $m_{X_{c}}$\mbox{(GeV)} ~~&~~ $m_{X_{b}}$\mbox{(GeV)}  \\
\hline
$0^{++}$      & $J_1$               &  $6.44\pm0.15$                  & $18.45\pm0.15$  \\
              & $J_2$               &  $6.59\pm0.17$                & $18.59\pm0.17$        \\
                & $J_3$               &  $6.47\pm0.16$                &$18.49\pm0.16$     \\
                    & $J_4$               &  $6.46\pm0.16$                 &$18.46\pm0.14$       \\
                       & $J_5$               &  $6.82\pm0.18$                  & $19.64\pm0.14$
\vspace{5pt}\\
$0^{-+}$        & $J_1^+$            &  $6.84\pm0.18$                 & $18.77\pm0.18$ \\
                     & $J_2^+$            &  $6.85\pm0.18$                 & $18.79\pm0.18$
\vspace{5pt}\\
$0^{--}$       & $J_1^-$              &  $6.84\pm0.18$                & $18.77\pm0.18$
\vspace{5pt}\\
$1^{++}$      & $J_{1\mu}^+$    & $6.40\pm0.19$                & $18.33\pm0.17$  \\
                    & $J_{2\mu}^+$    & $6.34\pm0.19$                & $18.32\pm0.18$
\vspace{5pt}\\
$1^{+-}$       & $J_{1\mu}^-$     & $6.37\pm0.18$                & $18.32\pm0.17$  \\
                    & $J_{2\mu}^+$    & $6.51\pm0.15$                & $18.54\pm0.15$
\vspace{5pt}\\
$1^{-+}$       & $J_{1\mu}^+$    & $6.84\pm0.18$                 & $18.80\pm0.18$ \\
                    & $J_{2\mu}^+$    & $6.88\pm0.18$               & $18.83\pm0.18$
\vspace{5pt}\\
$1^{--}$        & $J_{1\mu}^-$     & $6.84\pm0.18$                & $18.77\pm0.18$  \\
                    & $J_{2\mu}^-$     & $6.83\pm0.18$              & $18.77\pm0.16$
\vspace{5pt}\\
$2^{++}$       & $J_{1\mu\nu}$  & $6.51\pm0.15$                & $18.53\pm0.15$  \\
                     & $J_{2\mu\nu}$  & $6.37\pm0.19$                & $18.32\pm0.17$  \\
\hline\hline
\end{tabular}
\end{table}

The upper limit of $n$ can be obtained by studying the convergence of the OPE series, by requiring the perturbative term to be larger than the gluon condensate term. The upper bound $n_{max}$ usually increases with respect to the value of $\xi$. In principle, one can arrive at a region in the $(n, \xi)$ plane where the lowest lying resonance dominates the moments in Eq.~(\ref{moments}) and the OPE series has good convergence. We shall perform our analysis in such $(n, \xi)$ parameter regions to establish reliable moment sum rules for hadron masses. 

We show the numerical analysis for the fully-charm $cc\bar c\bar c$ tetraquark state with $J^{PC}=0^{-+}$ as an example, by using the interpolating current $J_1^+(x)$ for such quantum numbers in Ref.~\cite{Chen:2016jxd}. Considering the above criteria, we find that the upper bound $n_{max}$ will increase with respect to the value of 
$\xi=2.0, 3.0, 4.0, 5.0$. The mass curves are shown as a function of $n$ with different value of $\xi$ in Fig.~\ref{cccc0-+}, in which the mass plateaus are obtained to extract the hadron mass as 
\begin{align}
m_{cc\bar c\bar c}=6.84\pm0.18~\mbox{GeV},
\end{align}
in which the error comes from the uncertainties of the parameters $(n, \xi)$, gluon condensate and heavy quark mass. After carefully numerical analyses, we obtain systematical mass spectra for the $cc\bar c\bar c/bb\bar b\bar b$, $bc\bar b\bar c$ and $bb\bar c\bar c$ tetraquark states in Table~\ref{mass:QQQQ}, Table~\ref{mass:bcbc} and Table~\ref{mass:ccbb}, respectively.

In Table~\ref{mass:QQQQ}, one notes that the masses for the fully-charm $cc\bar c\bar c$ tetraquark states with $J^{PC}=0^{-+}, 1^{-+}$ are predicted to be 6.8-6.9 GeV, which are in good agreement with the mass of $X(6900)$ 
observed by LHCb~\cite{LHCb:2020bwg}. Moreover, the fully-charm $cc\bar c\bar c$ tetraquarks with $J^{PC}=0^{++}, 2^{++}$ are predicted around 6.3-6.6 GeV, which are consistent with the mass region of the broad structure in the 
di-$J/\psi$ spectrum. Such results successfully predicted the existence of these di-$J/\psi$ structures before the LHCb's observation. For the fully-bottom $bb\bar b\bar b$ tetraquark systems, our results show that they are below the two-bottomonium thresholds, which shall restrict such two-body strong decays. 

\begin{table}[h!]
\renewcommand\arraystretch{1.2} 
\caption{The mass spectra of $ bc\bar{b}\bar{c}$ tetraquark states with various $J^{PC}$.}\label{mass:bcbc}
\vspace{6pt}
\begin{tabular}{c c c c c}
  \hline
  \hline
 $J^{PC}$ & $Current$ & Mass$[\mathrm{GeV}]$ &   $Current$ & Mass$[\mathrm{GeV}]$ \vspace{1ex}  \\
              &    $[\mathbf{\bar{3}\otimes3}]_c$   &                                       &     $[\mathbf{6\otimes\bar{6}}]_c$     &             \vspace{1ex}  \\
  \hline
$0^{++}$ & $J_1$ & $12.28^{+0.15}_{-0.14}$ & $j_1$ & $12.37^{+0.15}_{-0.14}$ \vspace{1ex}  \\
           & $J_2$ & $12.46^{+0.17}_{-0.15}$ & $j_2$ & $12.29^{+0.15}_{-0.12}$ \vspace{1ex}  \\
           & $J_{5\mu\nu}$ & $12.35^{+0.14}_{-0.12}$ & $j_{5\mu\nu}$ & $12.32^{+0.15}_{-0.12}$ \vspace{1ex}  \\
           & $J_{5\mu\nu}(T)$ & $12.45^{+0.17}_{-0.15}$  & $j_{5\mu\nu}(T)$ & $12.29^{+0.14}_{-0.12}$ \vspace{1ex}  
           \vspace{6pt}\\
$0^{-+}$ & $J_{3\mu}$ & $12.99^{+0.22}_{-0.18}$ & $j_{3\mu}$ & $13.16^{+0.23}_{-0.20}$ \vspace{1ex} \vspace{6pt} \\
$0^{--}$ & $J_{4\mu}$ & $12.98^{+0.22}_{-0.18}$ & $j_{4\mu}$ & $13.17^{+0.23}_{-0.19}$ \vspace{1ex}  \vspace{6pt}\\        
$1^{++}$ & $J_{3\mu}$ & $12.30^{+0.15}_{-0.14}$ & $j_{3\mu}$ & $12.36^{+0.16}_{-0.14}$ \vspace{1ex}  \vspace{6pt}\\
$1^{+-}$ & $J_{4\mu}$ & $12.32^{+0.15}_{-0.13}$ & $j_{4\mu}$ & $12.34^{+0.15}_{-0.14}$ \vspace{1ex}  \vspace{6pt}\\
              & $J_{6\mu\nu}(A)$ & $12.38^{+0.13}_{-0.12}$ & $j_{6\mu\nu}(A)$ & $12.30^{+0.14}_{-0.12}$ \vspace{1ex}  \vspace{6pt}\\
$1^{-+}$ & $J_{5\mu\nu}$ & $13.23^{+0.24}_{-0.20}$ & $j_{5\mu\nu}$ & $13.17^{+0.23}_{-0.20}$ \vspace{1ex} \vspace{6pt} \\
$1^{--}$ & $J_{6\mu\nu}$ & $12.91^{+0.19}_{-0.16}$ & $j_{6\mu\nu}$ & $13.13^{+0.22}_{-0.19}$ \vspace{1ex}  \vspace{6pt}\\          
$2^{++}$ & $J_{5\mu\nu}$ & $12.30^{+0.15}_{-0.14}$ & $j_{5\mu\nu}$ & $12.35^{+0.15}_{-0.14}$ \vspace{1ex}  \vspace{6pt}\\
  \hline
  \hline
\end{tabular}
\end{table}

\section{Summary}
In this talk, I have presented our investigations of the mass spectra for the $cc\bar c\bar c/bb\bar b\bar b$, $bc\bar b\bar c$ and $bb\bar c\bar c$ tetraquark states in QCD moment sum rule method. Our results strongly suggest that 
the $X(6900)$ resonance to be a $cc\bar c\bar c$ tetraquark state with $J^{PC}=0^{-+}$ or $1^{-+}$, while the broad structure in LHCb's observation to be a $cc\bar c\bar c$ tetraquark state with $J^{PC}=0^{++}$ or $2^{++}$. For the fully-bottom $bb\bar{b}\bar{b}$ tetraquark states in all channels, their masses have been predicted to be lower than the $\eta_b\eta_b$ and $\Upsilon(1S)\Upsilon(1S)$ thresholds, indicating that they are not able to decay into the two bottomonium final states via strong interaction. 

The S-wave positive parity $bc\bar b\bar c$ and $cc\bar{b}\bar{b}$ tetraquark states have been predicted to be lower than the two-meson strong decay thresholds, implying that they are probably stable against the strong interaction. The P-wave negative parity $bc\bar b\bar c/cc\bar{b}\bar{b}$ tetraquarks can decay into the charmonium plus bottomonium and B-meson pair final states via the spontaneous dissociation mechanism. These tetraquark states may be observed at facilities such as LHCb, CMS and RHIC in the future.

\begin{table}
\begin{center}
\renewcommand{\arraystretch}{1.6}
\caption{The hadron mass predictions for the $cc\bar{b}\bar{b}$ tetraquark states with various $J^{P}$.}
\begin{tabular*}{8.4cm}{ccc|ccc}
 \hline  \hline      
   Current             & $J^{P} $     &Mass(\text{GeV)}             &Current          &$J^{P} $   & Mass(\text{GeV)}  \\ \hline
    $\eta_{1}^{+}$     & $0^{+}$     & $13.32_{-0.24}^{+0.30} $    & $\eta_{1}^{-}$      & $0^{-}$      &  $12.97_{-0.21}^{+0.25} $  \\              
   $\eta_{2}^{+}$      & $0^{+}$     & $12.41 _{-0.17}^{+0.21} $    & $\eta_{2}^{-}$      & $0^{-}$      &  $12.72_{-0.19}^{+0.22} $   \\

   $\eta_{3}^{+}$      & $0^{+}$     & $12.33 _{-0.15}^{+0.18} $    & $\eta_{3}^{-}$      & $0^{-}$      &  $13.16_{-0.24}^{+0.29} $   \\

   $\eta_{4}^{+}$      & $0^{+}$     & $12.36 _{-0.15}^{+0.18} $    & & &  \\  
   $\eta_{5}^{+}$      & $0^{+}$     & $12.36 _{-0.16}^{+0.19} $   &   &  &  \\  [.3cm]
    $\eta_{1\mu}^{+}$  & $1^{+}$     & $13.35_{-0.26}^{+0.33} $     & $\eta_{1\mu}^{-}$   & $1^{-}$      &  $13.02_{-0.21}^{+0.26} $   \\ 
    $\eta_{2\mu}^{+}$  & $1^{+}$     & $13.33_{-0.22}^{+0.28} $     & $\eta_{2\mu}^{-}$   & $1^{-}$      &  $12.77_{-0.19}^{+0.24} $     \\
    $\eta_{3\mu}^{+}$  & $1^{+}$     & $12.36_{-0.16}^{+0.19} $     & $\eta_{3\mu}^{-}$   & $1^{-}$      &  $12.99_{-0.22}^{+0.27} $    \\
    $\eta_{4\mu}^{+}$  & $1^{+}$     & $12.34_{-0.15}^{+0.18} $     & $\eta_{4\mu}^{-}$   & $1^{-}$      &  $12.87_{-0.20}^{+0.24} $     \\ [.3cm]
     $\eta_{1\mu\nu}^{+}$     & $2^{+}$                   &  $13.41_{-0.26}^{+0.34} $   &&&    \\
     $\eta_{2\mu\nu}^{+}$     & $2^{+}$                   &  $12.37_{-0.16}^{+0.19} $   &&&    \\
       \hline\hline  
\label{mass:ccbb}
\end{tabular*}
\end{center}
\end{table}

\section*{Acknowledgements\label{Ack}}
This work is supported in part by National Key R$\&$D Program of China under Contracts No. 2020YFA0406400, the National Natural Science Foundation of China under Grants No. 11722540, No. 11975033, No. 12075019, and No. 12070131001, the Fundamental Research Funds for the Central Universities.


\end{document}